\documentclass[12pt]{article}
\usepackage[cp866]{inputenc}
\usepackage[russian]{babel}
\usepackage[dvips]{graphicx}
\usepackage{graphicx}
\usepackage{epsfig}
\usepackage{amsmath}

\usepackage{caption}
\usepackage{subcaption}
\usepackage{amssymb}

\def\be{\begin{equation}}
\def\bea{\begin{eqnarray}}
\def\ee{\end{equation}}
\def\eea{\end{eqnarray}}

\def\t{\vartheta}
\def\p{\partial}

\newcommand{\jgr}{    {J. Geophys. Res.}}

\begin{document}

\begin{center}
{\Large \bf Hamiltonian for guiding center motion: symplectic structure approach}\\

\vskip 10mm

{\Large A. I. Neishtadt$^{1,2}$, A.~V.~Artemyev$^{3,2}$\\

\vskip 5mm

{\large \it $^{1}$ Department of Mathematical Sciences, \\ Loughborough University, UK}\\
{\large \it $^{2}$ Space Research Institute, Moscow, Russia}\\
{\large \it $^{3}$ Institute of Geophysics and Planetary Physics, University of California, Los Angeles, California, USA}\\
}
\end{center}

\section*{Abstract}
The guiding center approximation represents a very powerful tool for analyzing and modeling a
charged particle motion in strong magnetic fields. This approximation is based on conservation of the
adiabatic invariant, magnetic moment.  Hamiltonian equations  for the guiding centre motion are  traditionally intoduced  using
 a non-canonical symplectic structure. Such approach requires application of non-canonical Hamiltonian perturbation theory for calculations of the magnetic moment corrections. In this study we present an alternative approach with  canonical Hamiltonian equations for  guiding centre motion in time-dependent electromagnetic fields.   We show that the derived Hamiltonian
decouples three types of motion (gyrorotation, field-aligned motion, and across-field drifts), and each
type is described by a pair of conjugate variables. This form of Hamiltonian and symplectic structure
allows simple introduction of adiabatic invariants and can be useful for analysis of various plasma
systems.
\section{Introduction}
Charged particle dynamics in electromagnetic fields includes a Larmor (gyro) rotation around a magnetic field. For a sufficiently strong field, this rotation is the fastest motion in the system, and thus such fast periodic motion allows averaging
which gives the adiabatic invariant -- magnetic moment \cite{bookAlfven}. The guiding center approximation \cite{bookNorthrop63,bookSivukhin65} describes such averaged charged particle dynamics and represents a very powerful tool for investigation of various plasma systems (see, e.g., review in Ref. \cite{Cary&Brizard09} and references therein).

To derive equations describing guiding center motion of charged particles, one needs to introduce the magnetic moment as a new variable. Although, the required canonical transformation (through the generating function) has been proposed \cite{Gardner59}, this transformation has found almost no implementations in plasma physics (see discussion in Ref. \cite{Grebogi79}). Then, the alternative approach was proposed in Refs. \cite{Littlejohn79, Littlejohn83} which considers non-canonical variable transformations that result in the Hamiltonian equations with a non-standard (non-canonical) symplectic structure for the particle guiding center \cite{Tao07, Cary&Brizard09}. This is the most widespread approach applied for description of a guiding center motion in many space and laboratory plasma systems \cite{Boozer80, Chan98, Birn04:pop, Brizard&Hahm07, Ukhorskiy11}. The Hamiltonian derived using non-canonical variable transformation requires application of the non-canonical perturbation theory for evaluations of the higher order corrections. This complicates the usage of this Hamiltonian for description of plasma systems with a magnetic moment destruction (even if this destruction is weak), e.g., systems with the resonant wave-particle interaction violating the magnetic moment conservation.

Even in systems with the magnetic moment destruction due to magnetic field temporal/spatial inhomogeneities, the magnetic moment is often well conserved for almost entire particle orbit and experiences some variations only around spatially localized region. Therefore, using magnetic moment as a charged particle coordinate (in the velocity space) can significantly simplify the particle dynamics description. In this paper, we present a methodology  of derivation of canonical Hamiltonian equations for  guiding centre motion in time-dependent electromagnetic field introduced through the symplectic structure approach. For the case of time-independent magnetic field see, e.g.,  \cite{bookAKN06}, Sect. 6.4.1.

\section{Original equations}
Consider motion of a nonrelativistic charged particle (charge $e$, mass $m$) in an electromagnetic field with the magnetic component ${\bf B}({\bf {r}},t)=(B_x,B_y,B_z)$ and the electric component ${\bf E}({\bf r},t)=(E_x,E_y,E_z)$.   Here ${\bf r}=(x,y,z)$ is the position vector of the particle in  Cartesian coordinates $x,y,z$, and $t$ is time. Equations of motion of the particle are
\begin{equation}
\label{lorentz}
\dot {\bf r}={\bf v}, \quad m\dot {\bf v}=e{\bf E}+\frac{e}{c} {\bf v}\times  {\bf B}
\end {equation}
Denote ${\bf p}= m {\bf v}$.  Introduce time as a new phase variable $\t$: $\dot\t=1$.  Go to the extended phase space with additional  phase variables $\t, P_{\t}$.
Introduce the nondegenerate 2-form
\begin{eqnarray}
\label{Theta}
 \Theta_{\t}  = d{\bf p} \wedge d{\bf r}+ dP_{\t}  \wedge d\t  - \frac{e}{c}\sum\limits_{i < j} {\Gamma _{ij} } dr_i  \wedge dr_j
  - e{\bf E}d\t \wedge d{\bf r}
  \end{eqnarray}
where  $\Gamma_{12}=-B_z$, $\Gamma_{13}=B_y$, $\Gamma_{23}=-B_x$, $(r_1, r_2,r_3)=(x,y,z)$. Maxwell' s equations
$$
 \nabla\cdot {\bf B} =0, \,  \nabla\times {\bf E}= -\frac{1}{c}\frac{\p {\bf B}}{\p t}
$$
imply that the form $\Theta_{\t}$ is closed: $d\Theta_{\t}=0$.  Thus $\Theta_{\t}$ can be used as a symplectic structure. The direct calculation shows that Hamiltonian equations with this symplectic structure and Hamilton's function
\begin{equation}
\label{H_initial}
H_{\t}  = P_{\t}  + \frac{1}{2m}{\bf p}^2
\end{equation}
are equivalent to  equations (\ref{lorentz}) (for time independent magnetic field this calculation is contained, e.g., in \cite{nt}, Sect. 13.1).

\section{Guiding centre Hamiltonian}
Make variable transformation ${\bf r} \mapsto {\bf \eta}=(\alpha, \beta, R_\parallel)$, where $\beta$ and $\alpha$ are normalized Euler potentials \cite{Stern70}, i.e. ${\bf B}=B_0\nabla\beta\times\nabla\alpha$, and $R_\parallel$ is a field-aligned coordinate. Here $B_0={\rm const}$ is a typical magnetic field magnitude. This variable transformation is always possible locally due to the Darboux  theorem (see Ref. \cite{bookArnold89:mechanics}, Sect. 43.B).
This transformation is determined by a vector function ${\bf f}({\bf \eta},\t)$, i.e. ${\bf r}={\bf f}({\bf \eta},\t)$. Then
\begin{equation}
\label{G_term}
\frac{e}{c}\sum\limits_{i < j} {\Gamma _{ij} } dr_i  \wedge dr_j  = \frac{e}{c}B_0 d\alpha  \wedge d\beta
 - e\sum\limits_{i < j} {\Gamma _{ij} {\bf h}_{ij}} d{\bf \eta } \wedge d\t
\end{equation}
where
\begin{equation}
{\bf h}_{ij}=\frac{1}{c}\left({{\frac{{\partial f_i }}{{\partial \t}}\frac{{\partial f_j }}{{\partial {\bf \eta }}} - \frac{{\partial f_j }}{{\partial \t}}}\frac{{\partial f_i }}{{\partial {\bf \eta }}}}\right)
\label{eq06a}
\end{equation}
(note $\partial/\partial{\bf \eta}$ is the gradient in ${\bf \eta}=(\alpha,\beta,R_\parallel)$ space).


Denote
\begin{equation}
 \hat R = \frac{{\partial {\bf f}}}{{\partial {\bf \eta }}} = \left( {\begin{array}{*{20}c}
   {\partial _\alpha  f_1 } & {\partial _\beta  f_1 } & {\partial _{R_a } f_1 }  \\
   {\partial _\alpha  f_2 } & {\partial _\beta  f_2 } & {\partial _{R_a } f_2 }  \\
   {\partial _\alpha  f_3 } & {\partial _\beta  f_3 } & {\partial _{R_a } f_3 }  \\
\end{array}} \right)
\label{eq07}
\end{equation}
where $\partial_{\bf \alpha}=\partial/\partial {\bf \alpha}$, etc.

Let us consider the first term of symplectic structure (\ref{Theta}):
\begin{eqnarray}
d{\bf p} \wedge d{\bf r} &=& d\left( {{\bf p}\, d{\bf r}} \right) = d\left( {{\bf p} \, d{\bf f}({\bf \eta },\t)} \right)  = d\left( {{\bf p}\hat Rd{\bf \eta } + {\bf p}{\bf V}d\t} \right) \nonumber\\
  &=& d\left( {{\bf P}d{\bf \eta } + {\bf P}\hat R^{ - 1} {\bf V}d\t} \right)
  = d{\bf P} \wedge d{\bf \eta } + dF \wedge d\t \label{eq08}
\end{eqnarray}
where ${\bf P}={\bf p}\hat R=(P_\alpha, P_\beta, P_\parallel)$ is a new momentum conjugate to ${\bf \eta}$, ${\bf V}=\partial{\bf f}/\partial \t$, and $F={\bf P}\hat R^{-1}{\bf V}$ (note vectors are considered to be vector-row or vector-column depending on equation meaning).

The last term of symplectic structure (\ref{Theta}) is
\begin{equation}
\label{E_term}
- e{\bf E}d\t \wedge d{\bf r} = - e{\bf E}d\t \wedge \hat Rd{\bf \eta }
\end{equation}
Thus, sum of Eq. (\ref{E_term}) and the last term of Eq. (\ref{G_term}) gives
\begin{eqnarray}
\label{tilde_E}
  - e\sum\limits_{i < j} {\Gamma _{ij} {\bf h}_{ij} } d{\bf \eta } \wedge d\t  - e{\bf E}d\t \wedge \hat Rd{\bf \eta }
  =  - e {\bf \tilde E}d\t \wedge d{\bf \eta }
   \end{eqnarray}

Using new variables $({\bf \eta}, {\bf P})$, we rewrite Hamilton's function (\ref{H_initial}) and symplectic structure (\ref{Theta}) as
\begin{eqnarray}
\label{H_Theta_1}
H_{\t}  &=& P_{\t}  + \frac{1}{{2m}}\left( {{\bf P}\hat R^{ - 1} } \right)^2 \\
 \Theta_{\t}  &=& d{\bf P} \wedge d{\bf \eta } + d\left( {F + P_{\t} } \right) \wedge d\t - \frac{e}{c}B_0 d\alpha  \wedge d\beta   - e {\bf \tilde E}d\t \wedge d{\bf \eta }
  \nonumber
\end{eqnarray}
We know that the form $\Theta_{\t}$ is closed:  $d\Theta_{\t}=0$. The terms   $d{\bf P}\wedge d{\bf \eta}$, $dP_{\t}\wedge d\t$,  $d\alpha\wedge d\beta$, and ${dF \wedge d\t}$ in Eq. (\ref{H_Theta_1}) are also closed forms. Thus we can write
\begin{equation}
d\left(   {\bf \tilde E} d\t  \wedge d{\bf \eta }  \right) = 0
\label{eq12}\nonumber
\end{equation}
which implies
\begin{eqnarray*}
d\left(   {\bf \tilde E} d\t  \wedge d{\bf \eta }  \right)&=& \left(- {\frac{{\partial \tilde E_\alpha  }}{{\partial \beta }} + \frac{{\partial \tilde E_\beta  }}{{\partial \alpha }}} \right)d\beta  \wedge d\alpha  \wedge d\t
 + \left( -{\frac{{\partial \tilde E_\alpha  }}{{\partial R_{\parallel} }} + \frac{{\partial \tilde E_\parallel  }}{{\partial \alpha }}} \right)dR_\parallel   \wedge d\alpha  \wedge d\t \nonumber\\
  &+& \left( -{\frac{{\partial \tilde E_\beta  }}{{\partial R_{\parallel} }} + \frac{{\partial \tilde E_\parallel  }}{{\partial \beta }}} \right)dR_\parallel   \wedge d\beta  \wedge d\t = 0 \label{eq13}
 \end{eqnarray*}
Therefore  electric field $\tilde{\bf E}$ is potential, i.e. the variable transformation ${\bf r}\mapsto{\bf \eta}$ results in vanishing of nonpotential part of electric field. We introduce the potential $\varphi$ for $\tilde{\bf E}$:  $\tilde{\bf E}=-\partial_{\bf \eta}\varphi$.  Hamilton's function and  symplectic structure take the forms
\begin{eqnarray}
H_{\t}  &=& P_{\t}  + \frac{1}{{2m}}\left( {{\bf P}\hat R^{ - 1} } \right)^2   \label{eq14}\\
 \Theta_{\t}  &=& d{\bf P} \wedge d{\bf \eta } + d\left( {F + P_{\t}  - e\varphi } \right) \wedge d\t - \frac{e}{c}B_0 d\alpha  \wedge d\beta  \nonumber
\end{eqnarray}
We return to the time-dependent system (time is not a phase variable anymore) and substitute $F$. Then we get the Hamiltonian system with the following Hamilton's function and symplectic structure:
\begin{eqnarray}
 H &=& \frac{1}{{2m}}\left( {{\bf P}\hat R^{ - 1} } \right)^2  - {\bf P}\hat R^{ - 1}{\bf V} + e\varphi \nonumber \\
 \Theta & =& d{\bf P} \wedge d{\bf \eta } - \frac{e}{c}B_0 d\alpha  \wedge d\beta  \label{eq15}
\end{eqnarray}
Hamiltonian (\ref{eq15}) can be rewritten as
\begin{eqnarray}
H &=& \frac{1}{{2m}}\left( {{\bf P}\hat R^{ - 1}  - m{\bf V}^{T}} \right)^2  + e\varphi  - \frac{1}{2}m{\bf V}^2  \label{eq16}
\end{eqnarray}
Here and below the superscript $T$  denotes the transposition. Terms in Hamiltonian $H$ from Eqs. (\ref{eq16}) can be formally separated to the kinetic energy (the first term containing ${\bf P}$) and the potential energy (the second term $e\varphi-m{\bf V}^2/2$).


In a strong magnetic field (if $B_0$ is large), the gyrorotation is the fastest type of motion. We  introduce new variables  (so-called guiding center variables): $\alpha=R_{\alpha} +cP_\beta/eB_0$, $\beta=R_{\beta}-cP_\alpha/eB_0$. Substituting $d\alpha=dR_{\alpha}+cdP_\beta/eB_0$, $d\beta=dR_{\beta}-cdP_\alpha/eB_0$ into Eq. (\ref{eq16}), we obtain a new symplectic structure:
\begin{equation}
\Theta  = \frac{c}{{eB_0 }}dP_\alpha   \wedge dP_\beta   + dP_\parallel   \wedge dR_\parallel   + \frac{{eB_0 }}{c}dR_\beta   \wedge dR_\alpha
\label{eq17}
\end{equation}
Symplectic structure (\ref{eq17}) has a standard (canonical) form ``$d{\bf p}\wedge d{\bf r}$'' and defines three pairs of canonically conjugate variables: $(P_\alpha, cP_\beta/eB_0)$, $(P_\parallel, R_\parallel)$, $(R_\beta, eB_0 R_\alpha/c)$.  In Equation (\ref{eq17}) the first term relates to gyrorotation, the second term relates to a particle field-aligned motion, and the third term relates to a particle cross-field drift. Thus, three types of motion have been separated in the symplectic structure.

Let us consider the quadratic form of the kinetic energy in Hamiltonian (\ref{eq16}):
\begin{equation}
 \left( {{\bf P}\hat R^{ - 1} } \right)^2
  = \left(\hat R^{ - 1} \left( \hat R^{ - 1} \right)^T {\bf P} \right)\cdot{\bf P}
   = \left( {\hat N{\bf P}} \right) \cdot {\bf P}\label{eq17a}
 \end{equation}
where $\hat N$ is $3\times 3$ matrix, and index $T$ means the transposed matrix. We use this form $(\hat N {\bf P})\cdot {\bf P}$ for following Hamiltonian transformations (note in the first term of Eq.(\ref{eq17a}) $\bf P$ is considered as a vector-row, and in the second and third terms - as a vector-column, $\cdot$ denotes the standard  scalar product).

The kinetic energy in Hamiltonian (\ref{eq16}) is a second order polynomial of $P_\alpha$, $P_\beta$, $P_\parallel$. This polynomial has a minimum at some value of vector ${\bf P}_\bot=( P_\alpha, P_\beta)$,  ${\bf P}_\bot^*={\bf C}_0(R_\parallel, {\bf R}_\bot,t)P_\parallel+{\bf C}_1(R_\parallel, {\bf R}_\bot, t)$, where ${\bf R}_\bot=( R_\alpha, R_\beta)$. Thus, we can write:
\begin{eqnarray}
H &=& \frac{1}{{2m}}\left( {\hat M\left( {{\bf P}_ \bot   - {\bf P}_ \bot ^* } \right)} \right)\cdot \left( {{\bf P}_ \bot   - {\bf P}_ \bot ^* } \right) \nonumber\\
  &+& \frac{1}{{2m}}g(P_\parallel-P_\parallel^*)^2  + e\varphi  - \frac{1}{2}m{\bf V}^2  \label{eq18}\\
 \Theta  &=& \frac{c}{{eB_0 }}dP_\alpha   \wedge dP_\beta   + dP_\parallel   \wedge dR_\parallel   + \frac{{eB_0 }}{c}dR_\beta   \wedge dR_\alpha   \nonumber
 \end{eqnarray}
where $g=g(R_\parallel, {\bf R}_\bot,t)$, $P_\parallel^*=P_\parallel^*(R_\parallel, {\bf R}_\bot,t)$, and $2\times2$ matrix $\hat M=\hat M({R_\parallel, \bf R}_\bot,t)$ are defined by coefficients of $3\times3$ matrix $\hat N$ with ${\bf \eta}={\bf \eta}(R_\parallel, {\bf R}_\bot, c{\bf P}_\bot /eB_0,t)$. For strong magnetic field, $B_0$, $H$ can be expanded over $c{\bf P}_\bot /eB_0$: $H = H_0  + ({c}/({{eB_0 }}))H_1$ and
\begin{eqnarray}
 H_0  &=& \frac{1}{{2m}}\left( {\hat M_0 \left( {{\bf P}_ \bot   - {\bf P}_ \bot ^* } \right)} \right)\cdot \left( {{\bf P}_ \bot   - {\bf P}_ \bot ^* } \right)
   + \frac{1}{{2m}}g_0 (P_\parallel-P_\parallel^*)^2  + U_0 \nonumber \\
 U_0  &=& e\varphi _0  - \frac{1}{2}m{\bf V}_0^2  \label{eq19} \\
 \Theta & =& \frac{c}{{eB_0 }}dP_\alpha   \wedge dP_\beta   + dP_\parallel   \wedge dR_\parallel   + \frac{{eB_0 }}{c}dR_\beta   \wedge dR_\alpha   \nonumber
\end{eqnarray}
where  all functions with subindex $0$ depend on $(R_\parallel, {\bf R}_\bot, t)$. Hamiltonian equations for $H_0$ are
\begin{eqnarray}
\dot {\bf P}_\bot   &=& \Omega _0 \hat G \hat M_0 \left( {{\bf P}_ \bot   - {\bf P}_ \bot ^* } \right)\nonumber\\
 \dot R_\parallel   &=& \frac{{g_0 }}{m}\left(P_\parallel -P_\parallel^*\right) ,\quad \dot P_\parallel   =  - \frac{{\partial H_0 }}{{\partial R_\parallel  }} \label{eq20}\\
 \dot R_\alpha   &=& \frac{1}{{\Omega _0 m}}\frac{{\partial H_0 }}{{\partial R_\beta  }},\quad \dot R_\beta   = -\frac{1}{{\Omega _0 m}}\frac{{\partial H_0 }}{{\partial R_\alpha  }} \nonumber
 \end{eqnarray}
where $\Omega_0=eB_0/mc$, and
\begin{equation}
\hat G = \left( {\begin{array}{*{20}c}
   {0 } & {-1 }  \\
   {1 } & {0}  \\
\end{array}} \right)
\end{equation}
Equations (\ref{eq20}) show that ${\bf P}_\bot$ changes with time much faster than $(R_\parallel, P_\parallel)$ and ${\bf R}_\bot$ (note factor $\sim \Omega_0$ is large). This separation of time scales allows us to consider ${\bf P}_\bot$ motion with frozen $R_\parallel,  {\bf R}_\bot,t$ in the right hand side of the differential equation for ${\bf P}_\bot$. This motion is described  by the following Hamiltonian $H_{00}$ and symplectic structure $\Theta_0$:
\begin{equation} 
\label{harmonic}
H_{00}=  \frac{1}{{2m}}\left( {\hat M_0\left( {{\bf P}_ \bot   - {\bf P}_ \bot ^* } \right)} \right)\cdot \left( {{\bf P}_ \bot   - {\bf P}_ \bot ^* } \right), \quad 
\Theta_0=\frac{c}{{eB_0 }}dP_\alpha   \wedge dP_\beta 
\end{equation}
This is the motion of a linear oscillator with the frequency $\Omega=\Omega_0(\det \hat M_0)^{1/2}$. An important property of matrix $\hat M_0$  is that $(\det \hat M_0)^{1/2}=|{\bf B}|/B_0$. Thus  the frequency of ${\bf P}_\bot$ rotation is $\Omega=e|{\bf B}|/mc$. Denote $I,\theta$ action-angle variables for the Hamiltonian $H_{00}$ with the symplectic structure $dP_\alpha   \wedge dP_\beta $.  For the linear oscillator the action equals to the ratio of energy and frequency (see Ref. \cite{bookLL:mech}). Thus  $ H_{00} =(1/m)  (|{\bf B}|/B_0)I=(1/m)(\Omega/\Omega_0) I$,  $dP_\alpha   \wedge dP_\beta =dI \wedge d\theta$.   The  transformation of variables $(P_\alpha, P_\beta )\mapsto (I, \theta)$ is determined by a generating function  $S(P_\beta, I,R_\parallel ,P_\parallel, R_\alpha,  R_\beta, t)$. Here all variables but $P_\beta, I$ are considered as parameters.

\medskip
Consider canonical transformation of all variables 
$$
(P_\alpha, P_\parallel , R_\beta,  \frac{c}{{eB_0 }}P_\beta, R_\parallel,  \frac{eB_0}{{c }}R_\alpha  )\mapsto (\frac{c}{{eB_0 }}\tilde I,  \tilde P_\parallel,  \tilde R_\beta,   \tilde \theta,\tilde R_\parallel,   \frac{eB_0}{{c }}\tilde R_\alpha  )
$$ 
which is determined by the generating function
$$
 \frac{c}{{eB_0 }}S(P_\beta,\tilde  I,R_\parallel ,\tilde P_\parallel, R_\alpha,  \tilde R_\beta,t)+R_\parallel \tilde P_\parallel+\frac{eB_0}{{c}}R_\alpha \tilde R_\beta
$$
Old and new variables are related via formulas
\begin{eqnarray*}
\tilde \theta =\frac{\partial S}{\partial \tilde I}, \quad \tilde R_\parallel=R_\parallel +  \frac{c}{{eB_0 }}\frac{\partial S}{\partial \tilde P_\parallel},  \quad \tilde R_\alpha= R_\alpha+   \left(\frac{c}{{eB_0 }}\right)^2\frac{\partial S}{\partial \tilde R_\beta},\\
P_\alpha =\frac{\partial S}{\partial P_\beta}, \quad P_\parallel=\tilde P_\parallel + \frac{c}{{eB_0 }}\frac{\partial S}{\partial R_\parallel},  \quad  R_\beta=  \tilde R_\beta + \left(\frac{c}{{eB_0 }}\right)^2\frac{\partial S}{\partial R_\alpha},
\end{eqnarray*}
(tilde marks new variables).
So, the transformation of variables for all variables but $P_\alpha,P_\beta$ is close to identical. We expand the Hamiltonian with respect to differences of these new and old variables. 
The new Hamiltonian is $\tilde H= H+({c}/({{eB_0 }})){\partial S}/{\partial t}$.  The principal term $\tilde H_0$ of the new Hamiltonian  is the principal term $H_0$ of the old Hamiltonian in which $P_\alpha,P_\beta$ are expressed via $\tilde I, \tilde \theta$, and  old variables (without tildes) are replaced by new variables (with tildes).   Thus, the new Hamiltonian and symplectic structure have the form (tildes are omitted)
\begin{eqnarray*}
H&=&H_0+({c}/({{eB_0 }})) H_1 \nonumber \\
H_0  &=& (1/m)(\Omega/\Omega_0) I +\frac{1}{{2m}}g_0 (P_\parallel-P_\parallel^*) ^2  + U_0    \label{eq22}\\
\Theta  &=&  \frac{c}{{eB_0 }}dI \wedge d\theta  + dP_\parallel   \wedge dR_\parallel   + \frac{{eB_0 }}{c}dR_\beta  \wedge dR_\alpha   \nonumber
\end{eqnarray*}
Functions $g_0$, $\Omega$, and $U_0$ depend on $R_\parallel$, $R_{\alpha}, R_{\beta}$, and time. 

\medskip
  In the adiabatic approximation the action $I$ is an adiabatic invariant: $I={\rm const}$. 
  The magnetic moment of the particle  $\mu=H_{00}/B$ is  related to this adiabatic invariant  as follows: $\mu= I/(mB_0)$. 
In the adiabatic  approximation the leading order Hamiltonian $H_0$  describes the guiding center motion according to the following equations:
\begin{eqnarray*}
 \dot R_\parallel & =& \frac{{\partial H_0 }}{{\partial P_\parallel }} = \frac{{g_0 }}{m}\left( {P_\parallel  - P_\parallel^* } \right), \quad  \dot P_\parallel  =  - \frac{{\partial H_0 }}{{\partial R_\parallel }} \\
 \dot R_\alpha   &=& \frac{1}{{\Omega _0 m}}\frac{{\partial H_0 }}{{\partial R_\beta  }},\quad \dot R_\beta   =  - \frac{1}{{\Omega _0 m}}\frac{{\partial H_0 }}{{\partial R_\alpha  }}  \nonumber
 \end{eqnarray*}

\section{Discussion and conclusions}
We propose an approach  for introduction of a magnetic moment as a phase variable into equations of motion of a charged particle  in time-dependent electromagnetic field. The resulting equations have the form of a Hamiltonian system with the standard (canonical) symplectic structure. They describe dynamics of the magnetic moment as well as motion of the corresponding guiding centre. The final Hamiltonian describes separately the three types of a charged particle motion: the gyrorotation, the field-aligned motion, and cross-field drifts. This approach supplements the approach of the derivation of guiding center equation of motion in the form of Hamiltonian system with non-canonical symplectic structure  \cite{Littlejohn83,Cary&Brizard09}.

The proposed approach allows to use in the considered problem all results of canonical adiabatic perturbation theory (e.g. about exponential time and exponential accuracy of conservation of an adiabatic invariance in analytic one-frequency systems), see, e.g., \cite{bookAKN06}, Sect. 6.4 and references therein. It  allows also application of the canonical Hamiltonian perturbation procedure  for corrections to the guiding centre motion, because it provides not only the guiding center Hamiltonian $H_0$, but also the correction term $H_1$ (see Eq. (\ref{eq19})). This term is responsible for magnetic moment destruction.   If the magnetic field configuration contains some singularities (e.g., resonances where the gyrophase rate of change drops to zero), the proposed approach provides an estimate of the accuracy of magnetic moment conservation \cite{Neishtadt86, Cary86}.

\bibliographystyle{plain}

\end{document}